\documentclass{aa}
\usepackage[english]{babel}
\usepackage{txfonts}
\usepackage{graphicx}
\usepackage{epsf}
\usepackage{natbib,twoopt}
\usepackage[breaklinks=true]{hyperref} 
\bibpunct{(}{)}{;}{a}{}{,} 
\newcommandtwoopt{\citeads}[3][][]{\href{http://adsabs.harvard.edu/abs/#3}%
                                        {\citealp[#1][#2]{#3}}}
\newcommandtwoopt{\citepads}[3][][]{\href{http://adsabs.harvard.edu/abs/#3}%
                                        {\citep[#1][#2]{#3}}}
\newcommandtwoopt{\citetads}[3][][]{\href{http://adsabs.harvard.edu/abs/#3}%
                                        {\citet[#1][#2]{#3}}}
\newcommandtwoopt{\citeyearads}[3][][]%
   {\href{http://adsabs.harvard.edu/abs/#3}{\citeyear[#1][#2]{#3}}}
\begin{document}

\title{OH and H$_{\mathbf{2}}$O maser variations in W33B}

\author{P.~Colom\inst{1}
\and E.~E.~Lekht\inst{2} \and
M.~I.~Pashchenko\inst{2}
\and G.~M.~Rudnitskij\inst{2}
 }

\offprints{P.~Colom\protect\\
A complete version of Figure~7 is available in electronic form
from \url{http://www.aanda.org/}. OH and H$_2$O data in ASCII
format are available at the CDS via anonymous ftp at
cdsarc.u-strasbg.fr (130.79.128.5),
\url{http://cdsweb.u-strasbg.fr/cgi-bin/qcat?J/A+A/}, or
\url{http://comet.sai.msu.ru/~gmr/Maser_monitoring/W33B/}}

\institute{LESIA, Observatoire de Paris, Section de Meudon, CNRS,
UPMC, Universit\'e Paris-Diderot,
5 place Jules Janssen,\\ 92195 Meudon CEDEX, France\\
\email{Pierre.Colom@obspm.fr}
\and
Lomonosov Moscow State University, Sternberg Astronomical Institute,
13 Universitetskij prospekt, Moscow, 119234, Russia\\
\email{gmr@sai.msu.ru}
  }

\date{Received: 19 November 2013 / accepted 22 October 2014}

\abstract
   {The active star-forming region W33B is a source of OH
   and H$_2$O maser emission located in distinct zones around the
   central object.}
   {The aim was to obtain the complete Stokes pattern of
   polarised OH maser emission and to trace its variability and to
   investigate flares and long-term variability of the H$_2$O
   maser and evolution of individual emission features.}
   {Observations in the OH lines at a wavelength of 18~cm were
   carried out on the Nan\c{c}ay radio telescope (France) at a
   number of epochs in 2008--2014; H$_2$O line observations
   (long-term monitoring) at $\lambda = 1.35$~cm were performed
   on the 22-metre radio telescope of the Pushchino Radio
   Astronomy Observatory (Russia) between 1981 and 2014.}
   {We have observed strong variability of the emission features
   in the main 1665- and 1667-MHz OH lines as well as in the
   1612-MHz satellite line. Zeeman splitting has been detected in
   the 1665-MHz OH line at 62~km~s$^{-1}$ and in the 1667-MHz
   line at 62 and 64~km~s$^{-1}$. The magnetic field intensity was
   estimated to be from 2 to 3~mG. The H$_2$O emission features
   form filaments, chains with radial-velocity gradients, or more
   complicated structures including large-scale ones.}
   {Long-term observations of the hydroxyl maser in the W33B
   region have revealed narrowband polarised emission in the
   1612-MHz line with a double-peak profile characteristic of
   Type IIb circumstellar masers. The 30-year monitoring of the
   water-vapour maser in W33B showed several strong flares of the
   H$_2$O line. The observed radial-velocity drift of the H$_2$O
   emission features suggests propagation of an excitation wave
   in the masering medium with a gradient of radial velocities.
   In OH and H$_2$O masers some turbulent motions of
   material are inferred.}


\maketitle
\titlerunning{The source of maser emission W33B}
\authorrunning{Colom et al.}

\section{Introduction}

Early studies showed that the thermal radio source \object{W33}
consists of two HII regions: strong and compact
\object{G12.80$-$0.20} and a fainter extended one
\object{G12.68$-$0.18} \citepads[see
e.g. ][]{1970AuJPA..14....1G}. 
Subsequent observations by \citetads{1978A&A....65..307G} 
showed that the extended component
(\object{G12.80$-$0.18}) in \object{W33} consists of several
faint discrete features. For the stronger source
\citetads{1975ApL....16...29G} 
found from H109$\alpha$, H134$\alpha$, and H158$\alpha$ radio
recombination lines (RRL) radial velocities 35.8, 32.0 and
38.6~km~s$^{-1}$, respectively. With this velocity the preferable
kinematic distance to the W33 complex was believed to be 4.4~kpc
\citepads{1983ApJ...267..638H}. 
\citetads{2006ApJS..165..338Q} 
observed toward W33 RRL C91$\alpha$ and C92$\alpha$. The radial velocity
of the extended region from the H134$\alpha$ line is
58~km~s$^{-1}$ \citepads{1975ApL....16...29G}. 
The accepted model of the W33 region was an interstellar
cloud expanding at a velocity of $\sim$13~km~s$^{-1}$, thus
producing in its radio line spectra two Doppler components at
approximately 32 and 58~km~s$^{-1}$.

\citetads{2008A&A...486..191P} 
resolved the kinematic distance ambiguity for a
number of galactic HII regions, among them W33, using 21 cm
HI absorption spectra. For this purpose they employed a method
proposed by \citetads{2003ApJ...582..756K}. 
From the absence of an HI absorption
feature at the tangential-point velocity in the direction of W33
\citetads{2008A&A...486..191P} 
concluded that W33 is at the near kinematic distance. They
estimated its distance as 4.9--5.1~kpc with a probable error of
$\pm$0.6~kpc. Finally, \citetads{1989A&A...213..339F} 
give for W33B a somewhat larger distance of 6.4~kpc.

However, the latest measurement of the trigonometric parallax of
H$_2$O masers in W33B
\citepads{2013A&A...553A.117I} 
yields for this source a distance of $2.40_{-0.15}^{+0.17}$~kpc,
thus placing it in the Scutum spiral arm. It was shown that all
the maser  sources in the W33 region, together with some nearby
masers outlining the Scutum
Arm \citepads{2014ApJ...793...72S}, 
are interconnected and are at the same distance; the sources A
and B possess proper motions with respect to C and to their
central stars. The proper motions of the H$_2$O masers are
discussed below.

Toward W33 three OH and H$_2$O maser emission sources are
observed: W33\,A, B, and C. The W33\,A and B masers are arranged
symmetrically relative to W33C and are at an angular distance of
$7^{\,\prime}\!.5$ from it. Sources B and C are associated with
the HII regions \object{G12.68$-$0.18} and
\object{G12.80$-$0.20}, respectively. Source A is at the
periphery of the W33 region, near the faint feature
\object{G12.91$-$0.28} \citepads{1978A&A....65..307G}. 

Maser emission in the main OH lines 1665 and 1667~MHz was detected
toward W33B by \citetads{1968ApJS...15..131G}. 
It was observed in a velocity interval of 58--66~km~s$^{-1}$.
\citetads{1970AuJPh..23..363R} 
made measurements in the main lines as well as in the 1612- and
1720-MHz satellite OH lines; they managed to detect only the
main-line emission. The velocity coincidence of this HII region
and the OH maser source testifies to their physical association.

In 1978 \citetads{1980SvAL....6...58P} 
detected on the Nan\c{c}ay radio telescope thermal emission and
absorption in the satellite OH lines toward W33B coming from a
$\sim 7^\prime\times 7^\prime$ extended source (molecular cloud)
as well as weakly polarised emission from a pointlike source in
the 1612-MHz line. Observations in 1978 on the same radio
telescope showed that the main-line OH emission is strongly
polarised circularly. The emission (and absorption) in a velocity
interval of 30--40~km~s$^{-1}$ belongs to the sources W33C and A,
and in an interval of 55--65~km~s$^{-1}$ to the source W33B.

The H$_2$O maser emission was detected toward W33 by\linebreak
\citetads{1977A&AS...30..145G} 
at virtually the same radial velocities as the OH emission. An
exception is the emission at small negative velocities in
W33C. In contrast to OH, a stronger H$_2$O source in
this region is W33B. Subsequent observations\linebreak
\citepads[][and this
work]{1981ApJ...250..621J} 
confirmed this characteristic of the H$_2$O masers.

The W33B region also hosts a strong source of maser emission of
methanol (CH$_3$OH) in the $5_1\,-6_0\,A^+$ 6.67-GHz rotational
line \citepads{1991ApJ...380L..75M}. 
The line profile consists of two peaks at
$v_{\mathrm{LSR}}\sim 52$ and $58$~km~s$^{-1}$.

According to the VLA observations of \citeauthor{1989A&A...213..339F}
(\citeyearads{1989A&A...213..339F},
\citeyearads{1999A&AS..137...43F}) 
for most maser sources associated with star-forming regions, OH
and H$_2$O masers occur in small groups with a diameter of less
than 0.03~pc. They have a common source of energy, but are
physically located in distinct zones.

\section{Observations and data presentation}

We observed the W33B radio source in the 18 cm hydroxyl lines at
various epochs on the telescope of the Nan\c{c}ay Radio Astronomy
Station of the Paris--Meudon Observatory (France). The method of
observation and processing of data was presented by
\citetads{2010ARep...54..599S} 
and \citetads{2012ARep...56...45L}. 
At declination $\delta=0^\circ$ the telescope beam at a
wavelength of 18~cm is $3.\!^\prime 5\times 19^\prime$ in right
ascension and declination. The telescope sensitivity at $\lambda
= 18$~cm and $\delta=0^\circ$ is 1.4~K/Jy. The system noise
temperature of the helium-cooled front-end amplifiers is from 35
to 60~K depending on the observational conditions.

\begin{figure}
\resizebox{\hsize}{!}{\includegraphics{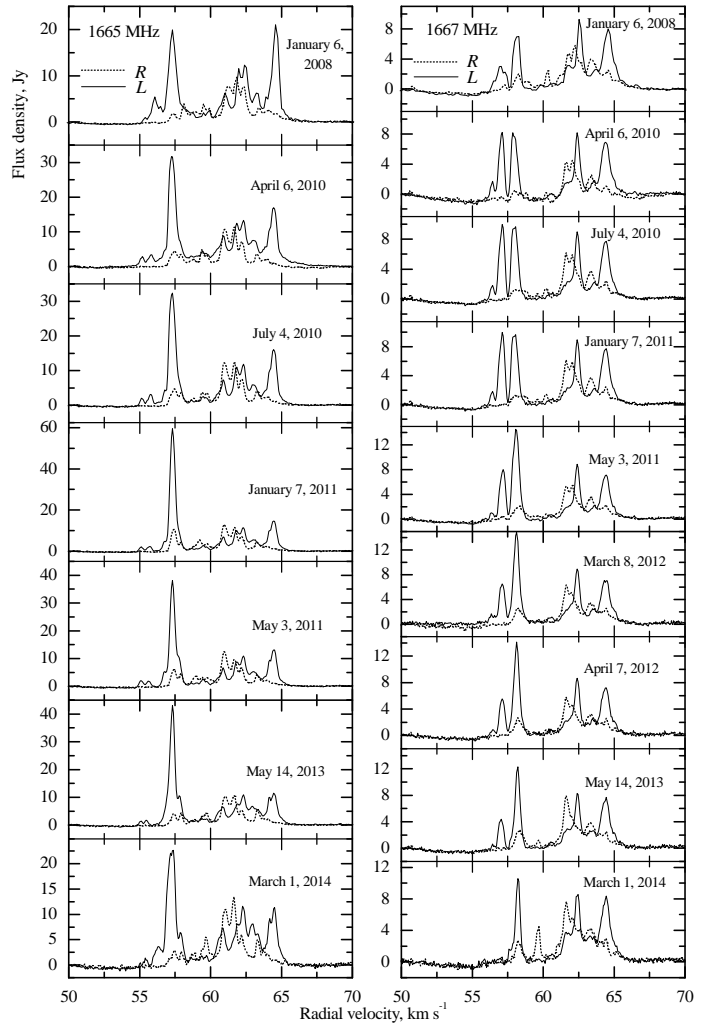}}
\caption{\small Spectra of OH maser emission
in the 1665- and 1667-MHz lines at various epochs. Solid curves:
left-hand circular polarisation; dashed: right-hand circular.}
\label{fig1}
\end{figure}

We observed H$_2$O maser emission in the 1.35~cm line toward W33B
($\alpha_{2000}=18^{\mathrm{h}}13^{\mathrm{m}}54.\!^{\mathrm{s}}7$,
$\delta_{2000}=-18^\circ 1^\prime 46.\!^{\prime\prime}5$) on the
22-metre radio telescope in Pushchino from February 1981 to
January 2014 with a half-power beamwidth of $2.\!^{\prime}6$. In
the observations of this source the system noise temperature with
a helium-cooled field-effect transistor (FET) front-end amplifier
was 120\,--\,270~K depending on the weather conditions. The
signal spectrum was measured by a 128-channel filter-bank
analyser with a velocity resolution of 0.101~km~s$^{-1}$, and
since the end of 2005 by a 2048-channel autocorrelator with a
resolution of 0.0822~km~s$^{-1}$. For a pointlike source an
antenna temperature of 1~K corresponds to a flux density of 25~Jy.

\begin{figure}
\resizebox{\hsize}{!}{\includegraphics{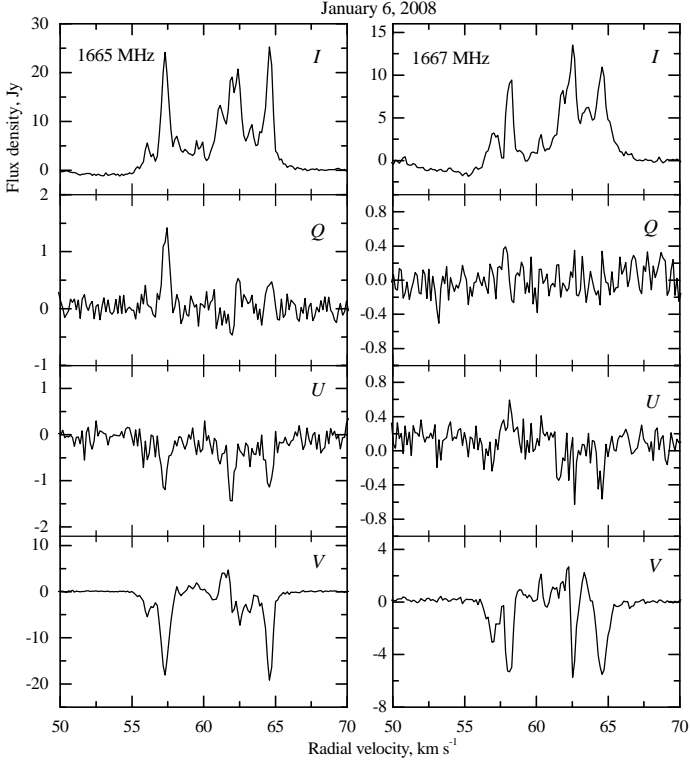}}
\caption{\small Stokes parameters of the main lines at 1665 and
1667~MHz for the epoch January 6, 2008.}
\label{fig2}
\end{figure}

Figure~\ref{fig1} presents the results of observations of hydroxyl
maser emission in the 1665- and 1667-MHz lines. The observations
in 2008 were carried out with a velocity resolution of
0.137~km~s$^{-1}$, and those of 2010--2014 with a resolution of
0.068~km~s$^{-1}$. The technique of the observations was
described in detail by
\citetads{2009ARep...53..541P} 
and \citetads{2010ARep...54..599S}. 

\begin{figure}
\resizebox{\hsize}{!}{\includegraphics{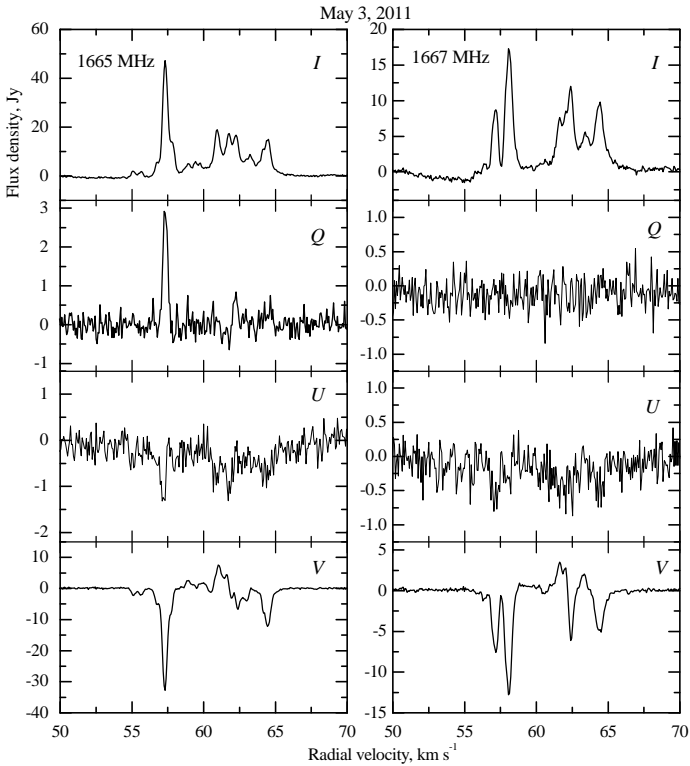}}
\caption{\small Same as in Fig.~\ref{fig2},
for the epoch May 3, 2011.} 
\label{fig3}
\end{figure}

\begin{figure}[!h]
\resizebox{\hsize}{!}{\includegraphics{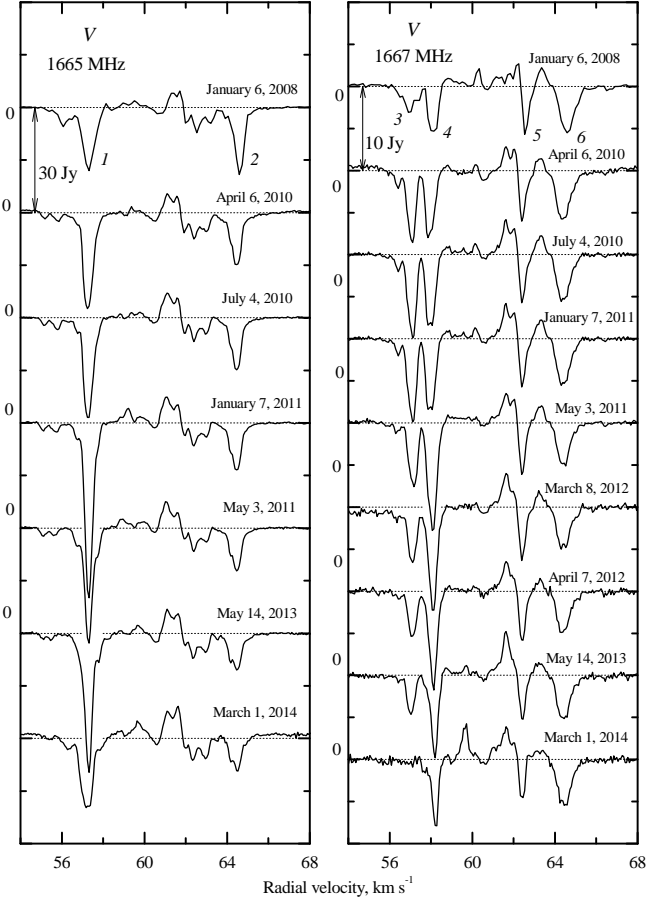}}
\caption{\small Central parts of the OH
main-line spectra for the Stokes parameter $V$ for various epochs
of the observations. Main spectral features are numbered.}
\label{fig4}
\end{figure}

\begin{figure}[!h]
\resizebox{\hsize}{!}{\includegraphics{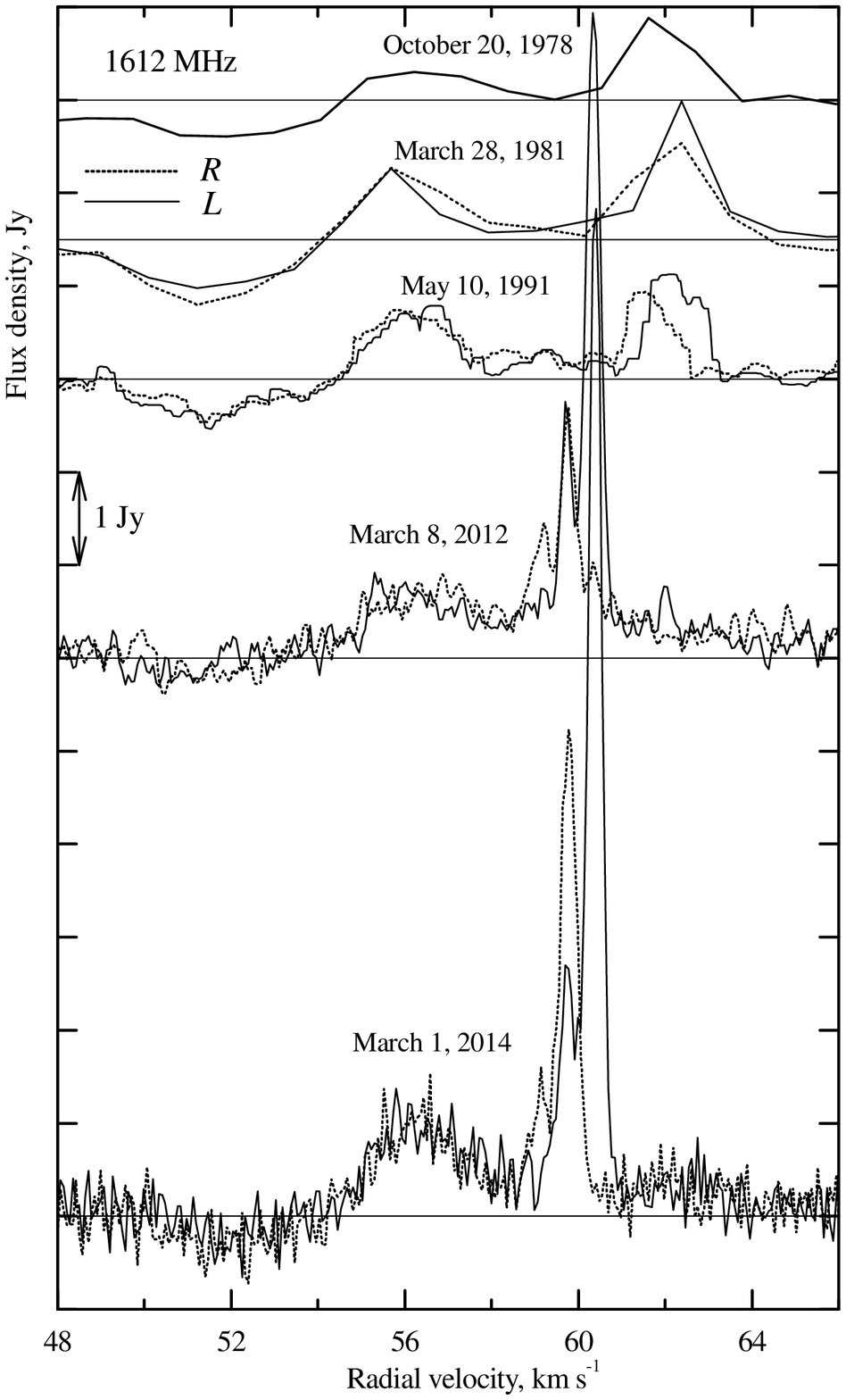}}
\caption{\small Spectra of OH maser emission in the satellite
line at 1612~MHz. Spectra in right- and left-hand circular
polarisation are shown with dotted and solid lines, respectively.
Strong narrowband polarised maser emission is visible in the 2012
and 2014 profiles.}
\label{fig5}
\end{figure}

\begin{figure}[!h]
\resizebox{\hsize}{!}{\includegraphics{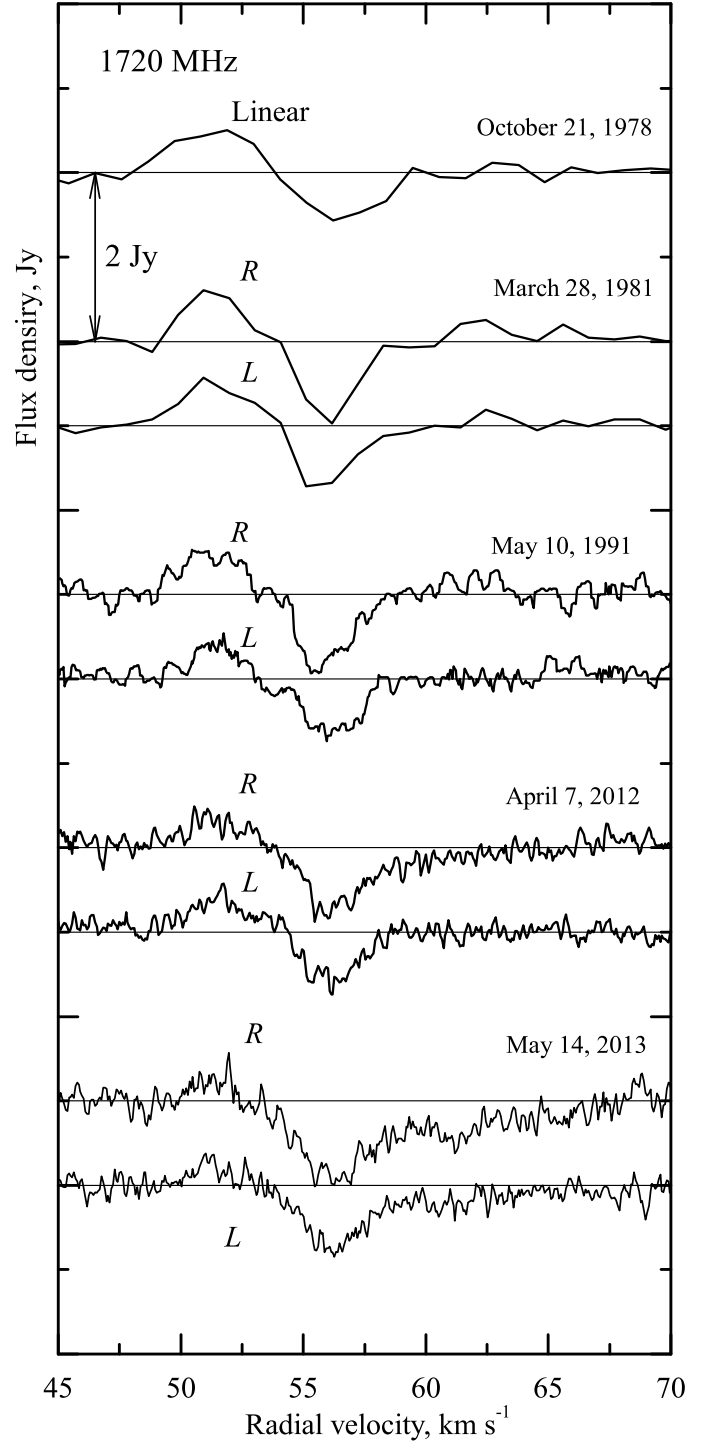}}
\caption{\small Spectra of OH maser emission in the satellite
line at 1720~MHz. There is no difference between right- (\emph{R})
and left-hand (\emph{L}) polarised profiles.} 
\label{fig6}
\end{figure}

Figures~\ref{fig2} and \ref{fig3} show Stokes parameters for the
main lines at 1665 and 1667~MHz at the epochs of January 6, 2008,
and May 3, 2011. Figure~\ref{fig4} presents central parts of the
spectra for Stokes parameter $V$ of the OH main lines at different
epochs. The main spectral features are numbered. The results of
our observations in the 1612- and 1720-MHz satellite lines are
presented in Figures~\ref{fig5} and \ref{fig6}, respectively.

Figure~\ref{fig7a} represents an atlas of the H$_2$O spectra for
the interval from November 1981 to January 2014. For technical
reasons, no observations were conducted between May 2006 and
December 2007. The horizontal axis is the velocity relative to
the local standard of rest (LSR). All the spectra are given in
the same radial-velocity scale. An arrow at the vertical axis
shows the scale in janskys. In the spectra zero baselines have
been drawn.

Superpositions of the H$_2$O spectra for various time intervals
(1--6) are shown in Figure~\ref{fig8}. The separation was done
according to the character of the spectral evolution. Averaged
spectra are shown with bold curves. The maser emission in the
velocity interval of 55\,--\,63~km~s$^{-1}$ is observed during
two time intervals (1 and 3). At the rest time the emission is
observed mainly in one or two narrow velocity intervals,
55\,--\,57.5 and 58\,--\,62~km~s$^{-1}$. From 2001 to 2012 the
velocity centroid of the averaged spectra of the main group
(58\,--\,62~km~s$^{-1}$) moved from 60.4 to 59.4~km~s$^{-1}$. In
addition, fluctuations of the velocity centroid calculated for
individual spectra were observed.

\section{Data analysis and discussion}

\subsection{Hydroxyl}
In 1978, using high angular resolution in right ascension of the
Nan\c{c}ay radio telescope, we searched for the
emission/absorption peaks in the satellite OH lines.
Figure~\ref{fig9} shows the emission/absorption intensity as a
function of right ascension for different spectral features.
Vertical arrows denote positions of radio continuum peaks
(W33C and B). Features~1 (62.3~km~s$^{-1}$) and 2
(56~km~s$^{-1}$) are unresolved, whereas feature~3
($\sim$50~km~s$^{-1}$) is extended. Peaks of features~1 and 2 in
the 1612-MHz line do not coincide; the former is detectable only
in the 1612-MHz line and can be associated with a maser spot.

Thus, according to our observations in 1978, 1991, 2012, and 2014,
(thermal) emission/absorption satellite-line features in the
velocity interval 49--58~km~s$^{-1}$ belong to an extended source,
whereas narrow emission features come from a pointlike source.

Of interest is broadband absorption in the main OH lines, which
probably covers the entire velocity interval filled by OH maser
emission features. It is visible at the $I$ profile edges and,
probably, in the $U$ Stokes profile (Figures~\ref{fig2} and
\ref{fig3}), thus suggesting a slight linear polarisation. The
velocity range can be assessed more certainly in the OH satellite
lines, especially in 1720 MHz, where OH is partly in thermal
emission and partly in absorption (see Figures~\ref{fig5} and
\ref{fig6}). It covers radial velocities $V_{\mathrm{LSR}}\approx
50\textrm{--}65$~km~s$^{-1}$, which probably represent the entire
velocity dispersion of material in the molecular cloud
surrounding the W33B maser.

\subsubsection{The structure of the OH maser source}
Figure~\ref{fig10}a shows the arrangement of the main-line OH
masers spots for the epoch 1991 \citepads{2000ApJS..129..159A}. 
Spots' radial velocities are indicated. The angular size of the
main-line masering region is 0.4$\times$0.4~arcsec; the regions of
emission in the 1665- and 1667-MHz lines are spatially separated.
The map centre $\Delta {\rm RA}=0$, $\Delta {\rm Dec}=0$
corresponds to RA(2000)=$18^{\rm h}13^{\rm m}54.75^{\rm s}$,
Dec(2000)=$-18\degr 1\arcmin 46.4\arcsec$. The distribution of
the maser spots is delineated with arcs (dashed curves). We
observe regular radial-velocity variations along the arc for the
1665-MHz maser spots. At first, the velocity decreases, then
increases. There are several clusters of maser spots with a small
$V_{\mathrm{LSR}}$ dispersion. An asterisk marks the presumed
location of the central star for the OH maser source based on the
arc shape i.e. on the large-scale structure of the source.
Another argument supporting our supposition is the large
separation between the OH and H$_2$O masers in W33B. In contrast
to \citetads{1989A&A...213..339F}, 
we think that each maser has its own source of energy.

\subsubsection{Main-line emission}
The OH maser in W33\,B is a typical representative of hydroxyl
masers associated with star-forming regions: the 1665-MHz line is
stronger than the 1667-MHz line.

Our main-line observations have shown that most OH emission
features in W33\,B have a high degree of circular polarisation
(see Figure~\ref{fig1}). Another peculiarity of the OH spectra is
strong variability of the most intense features. We have found no
Zeeman splitting for such features.

For the detection and study of Zeeman components we use Stokes
parameter $V$, the difference of the right- and left-hand
circular polarisations (see Figures~\ref{fig2}--\ref{fig4}). In
the 1665-MHz line the $\sigma$ Zeeman components can be a pair of
spectral features near 62~km~s$^{-1}$, and in the 1667-MHz line
these are pairs of features near 62 and 64~km~s$^{-1}$. The
splitting is 1.3, 0.7 and 1.1~km~s$^{-1}$. This corresponds to
intensities of the line-of-sight magnetic field of $B =-2.2$,
$-2.0$ and $-3.1$~mG, respectively ($B<0$ corresponds to the
field directed toward the observer).

It should be emphasised that in both OH main lines at
62~km~s$^{-1}$ we have obtained quite similar field intensities,
$-$2.2 and $-$2.0~mG. These features may be produced by the same
maser condensation.

As for previous measurements of the magnetic field toward W33\,B,
\citetads{1991ChJSS..11....1Z}, 
who observed W33B on the 43-m NRAO radio telescope in October
1981, found an average field $B=-5$~mG from the shifts of the mean
weighted velocities in the both main OH lines 1665 and 1667 MHz.
\citetads{2003ApJ...596..328F} 
found in their VLA observations of August 1991 several Zeeman
pairs of right- and left-hand circularly polarised features
within a velocity interval of 62.5--64.5~km~s$^{-1}$ in both OH
main lines. Their estimates for the field strength are from
$-$0.7 to $-$7.5~mG. The quoted estimates do not contradict ours,
both in the field direction and in the order of magnitude. The
direction of the magnetic field fits the general field pattern in
the Scutum Arm as directed counterclockwise when viewed from the
north galactic pole. Thus, during the collapse of the
protostellar gas cloud the general direction of the interstellar
field could have been conserved.

\subsubsection{Satellite-line emission}
As in the source W33C in the same region
\citepads{2012ARep...56..731C}, 
the profiles of the 1612- and 1720-MHz OH lines toward
\object{W33B} (Figures~\ref{fig5},~\ref{fig6}) consist of
emission and absorption components and mirror each other: peak
velocities are, respectively, 56.15 and 51.66~km~s$^{-1}$ in the
1612-MHz line, 51.47 and 56.23~km/s in the 1720-MHz line. The
mean difference is 4.6~km~s$^{-1}$ (for W33C it is
3.4~km~s$^{-1}$).

This structure is explained in a model of an OH source associated
with a molecular cloud where a maser is embedded in. If an IR
source is present in the cloud its radiation affects the
populations of the hyperfine structure sublevels of OH molecules
\citepads{1973SvA....16..597B}. 
In this model particulars of the IR pumping are such that
inversion of the 1720-MHz transition levels is accompanied by
anti-inversion in the 1612-MHz transition and vice versa. The
inversion or anti-inversion is determined by the angle between the
direction of propagation of the IR radiation and the direction of
local magnetic field. For a source embedded in the cloud one of
the satellites is inverted and the other one is anti-inverted; in
the other part of the cloud the situation is the opposite. This
effect was observed by us in W33C (Figure~\ref{fig9}).

In 2012 we detected narrowband 1612-MHz maser emission in both
circular polarisations. In 1991 this emission was absent. We
detected three emission features at radial velocities 59.1, 59.7
and 60.4~km~s$^{-1}$. Table~\ref{tab1} lists their flux densities
at different epochs. The columns labelled $F_{\textrm{R}}$ and
$F_{\textrm{L}}$ contain flux densities in the right- and
left-hand circular polarisations, respectively; the columns
labelled $p$ list degree of polarisation $p=\mid F_{\textrm{R}} -
F_{\textrm{L}}\mid\,/\,(F_{\textrm{R}} + F_{\textrm{L}})$\,.

\begin{figure}[!h]
\resizebox{\hsize}{!}{\includegraphics{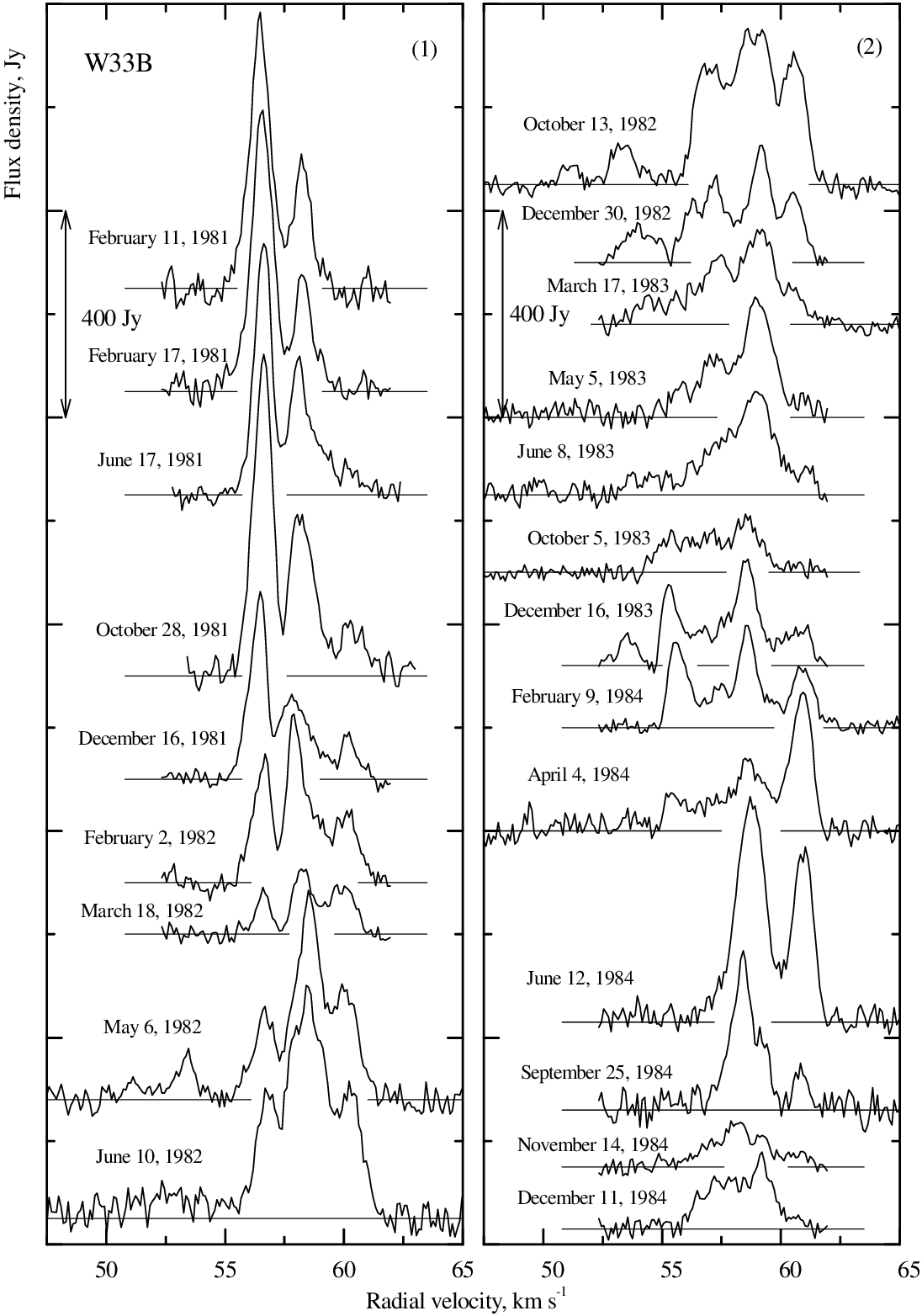}}
\caption{\small Spectra of the H$_2$O maser
emission in W33B.}
 \label{fig7}
\end{figure}

\addtocounter{figure}{-1}
\onlfig{
\begin{figure*}
\centering
\includegraphics[width=14cm]{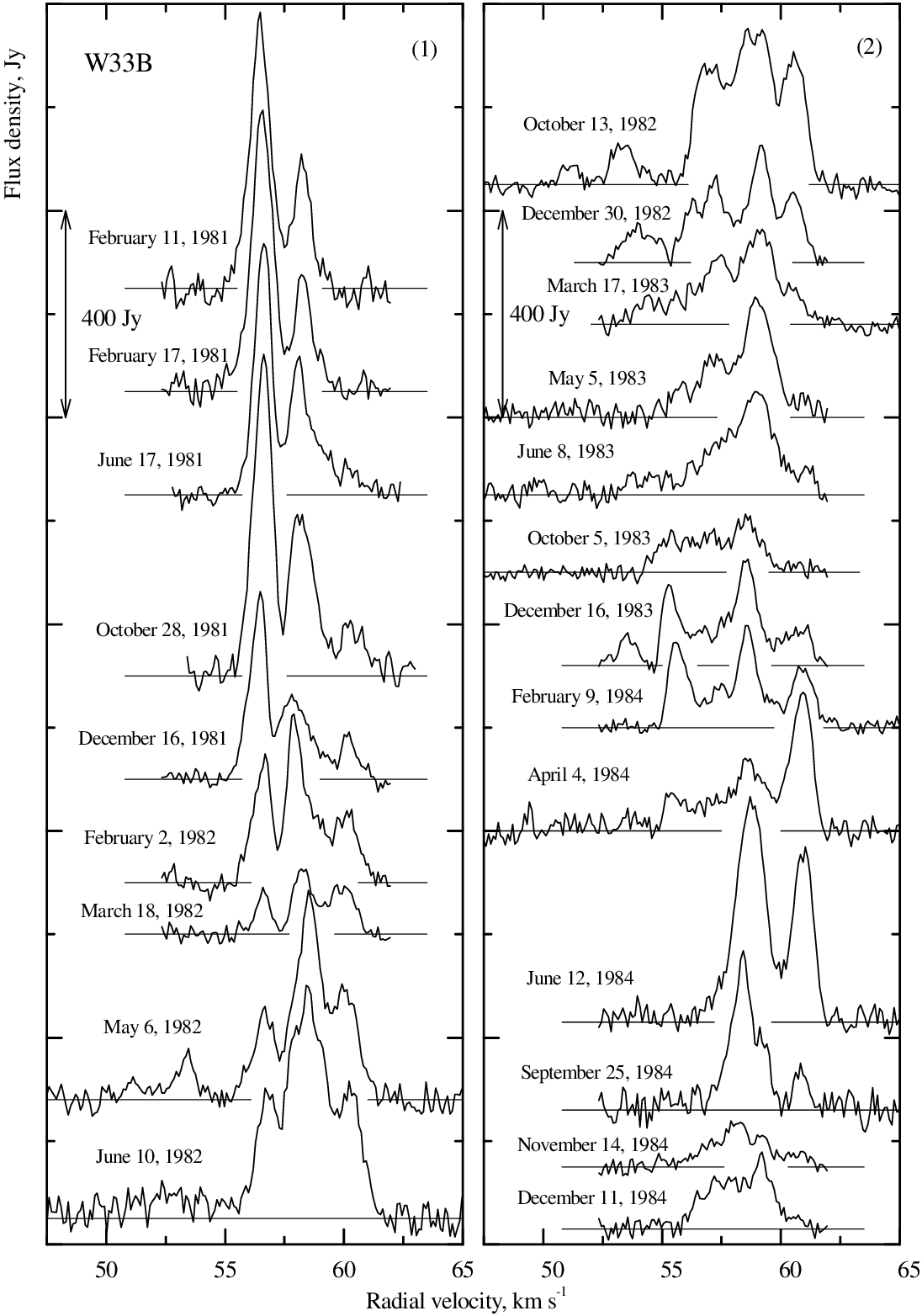}
\caption{\small Spectra of the H$_2$O maser
emission in W33B.}
\label{fig7a}
\end{figure*}
}

\addtocounter{figure}{-1}
\onlfig{
\begin{figure*}
\centering
    \includegraphics[width=14cm]{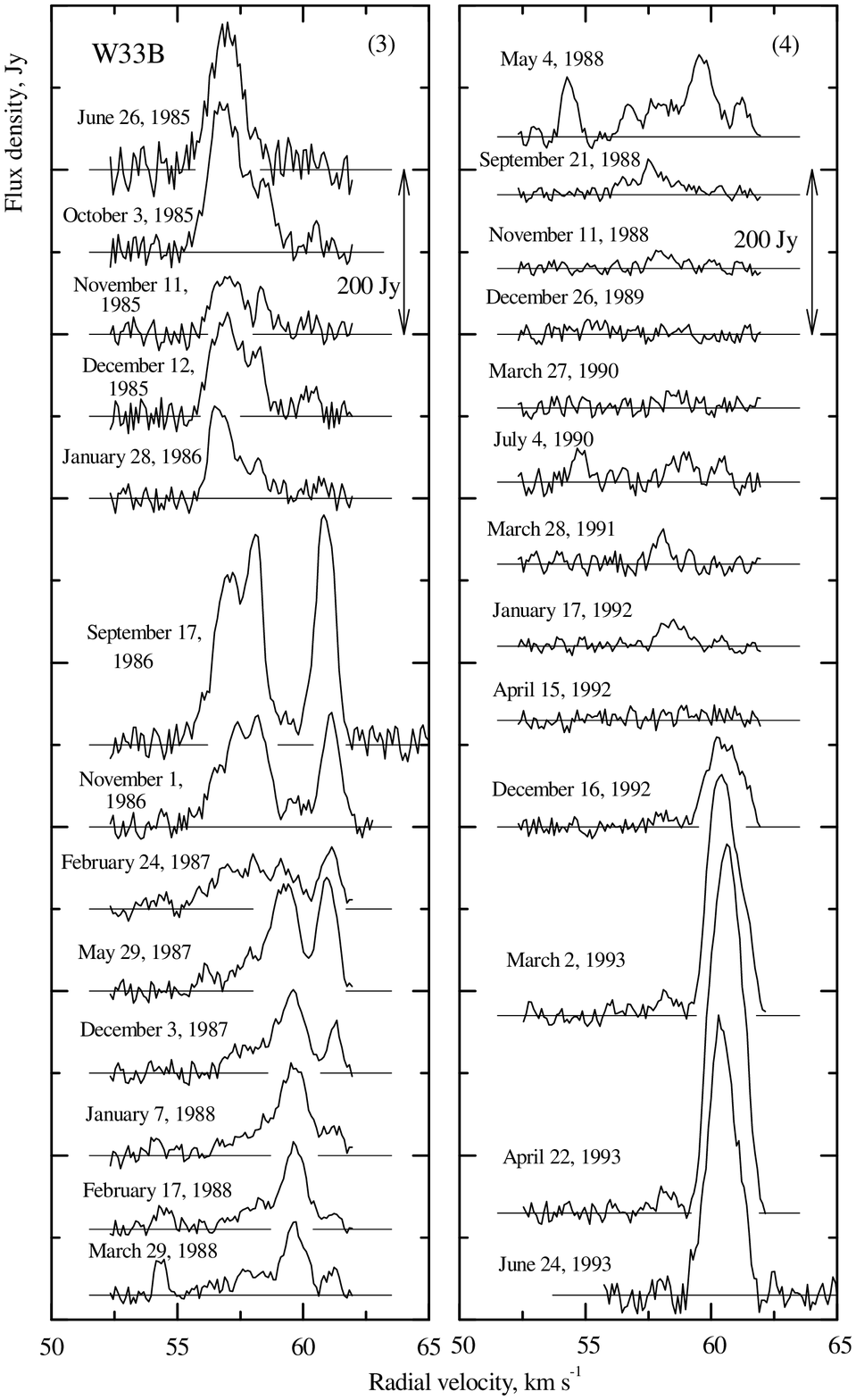} \caption{\small Continued.} \label{fig7b}
\end{figure*}
}

\addtocounter{figure}{-1}
\onlfig{
\begin{figure*}
\centering
    \includegraphics[width=14cm]{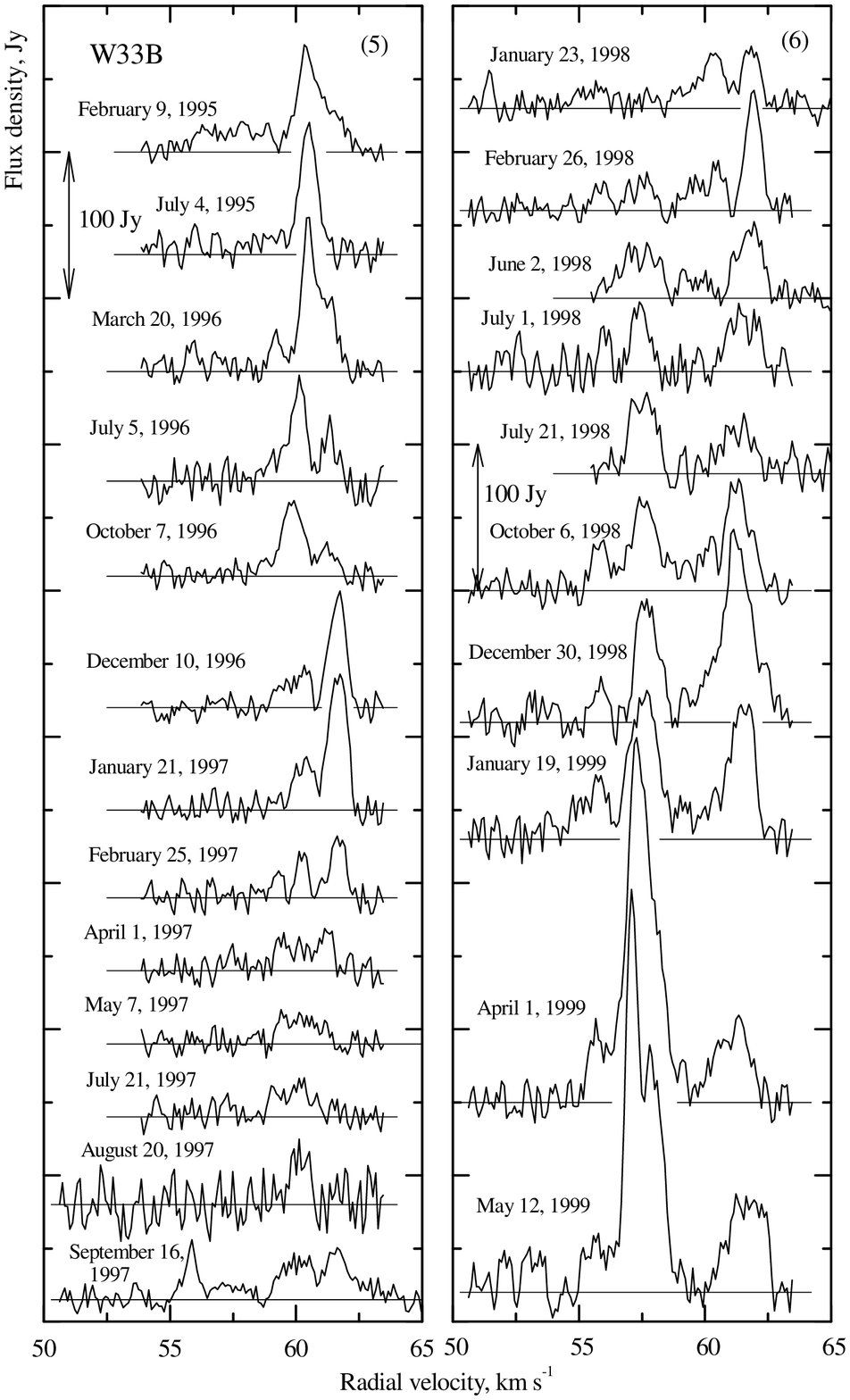} \caption{\small  Continued.} \label{fig7c}
\end{figure*}
}

\addtocounter{figure}{-1}
\onlfig{
\begin{figure*}
\centering
    \includegraphics[width=14cm]{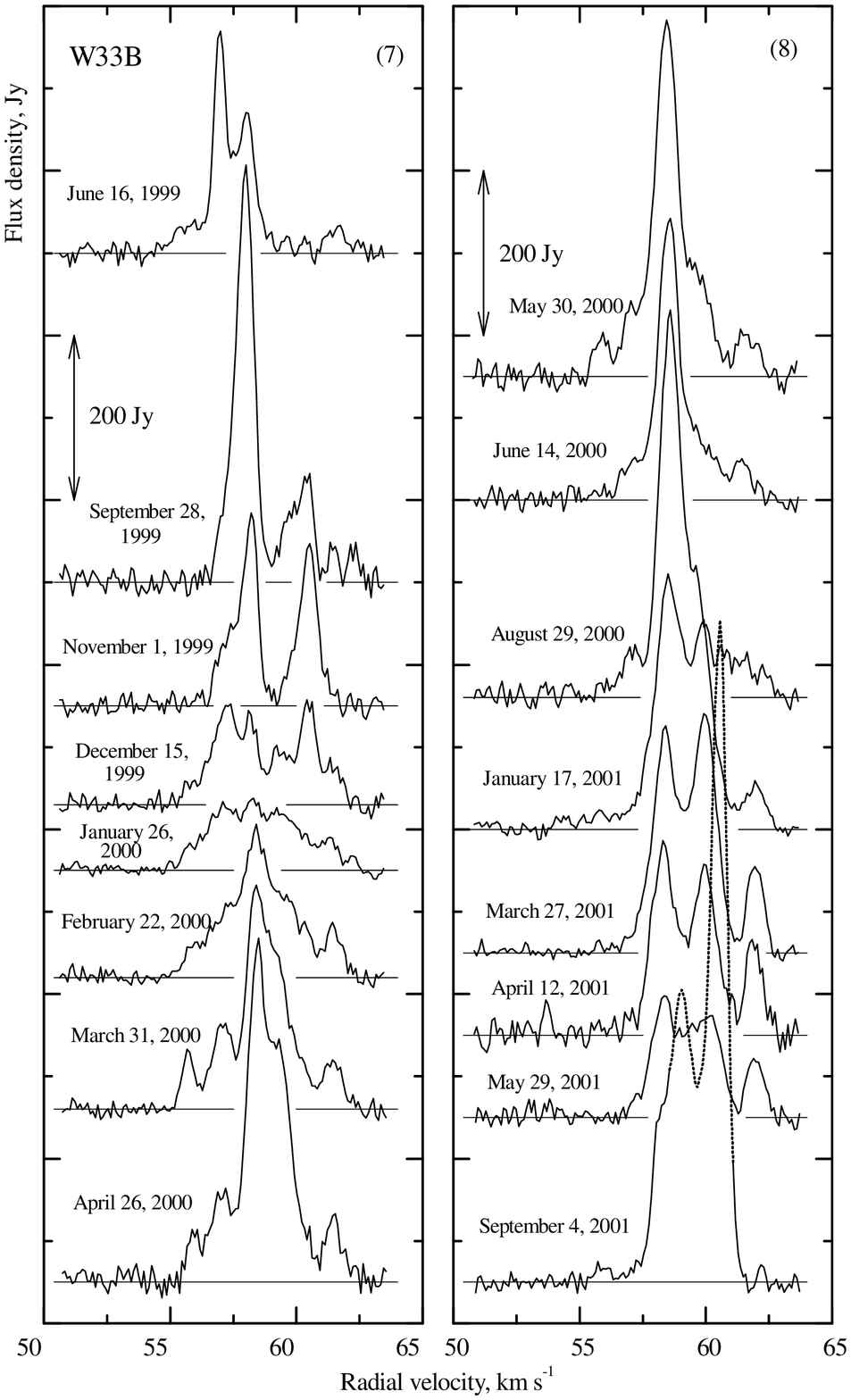} \caption{\small  Continued.} \label{fig7d}
\end{figure*}
}

\addtocounter{figure}{-1} \onlfig{
\begin{figure*}
\centering
    \includegraphics[width=14cm]{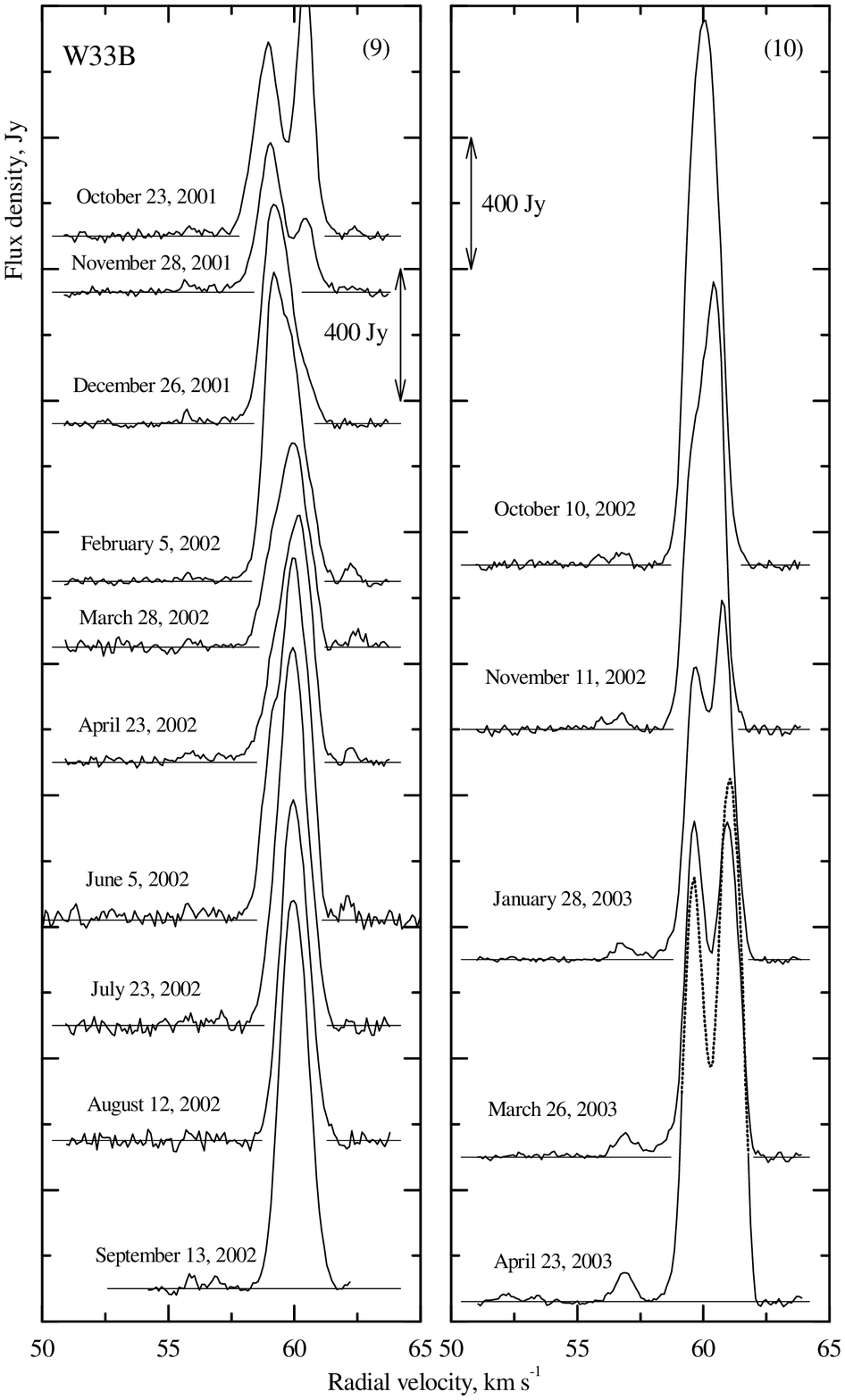} \caption{\small  Continued.} \label{fig7e}
\end{figure*}
}

\addtocounter{figure}{-1}
\onlfig{
\begin{figure*}
\centering
    \includegraphics[width=14cm]{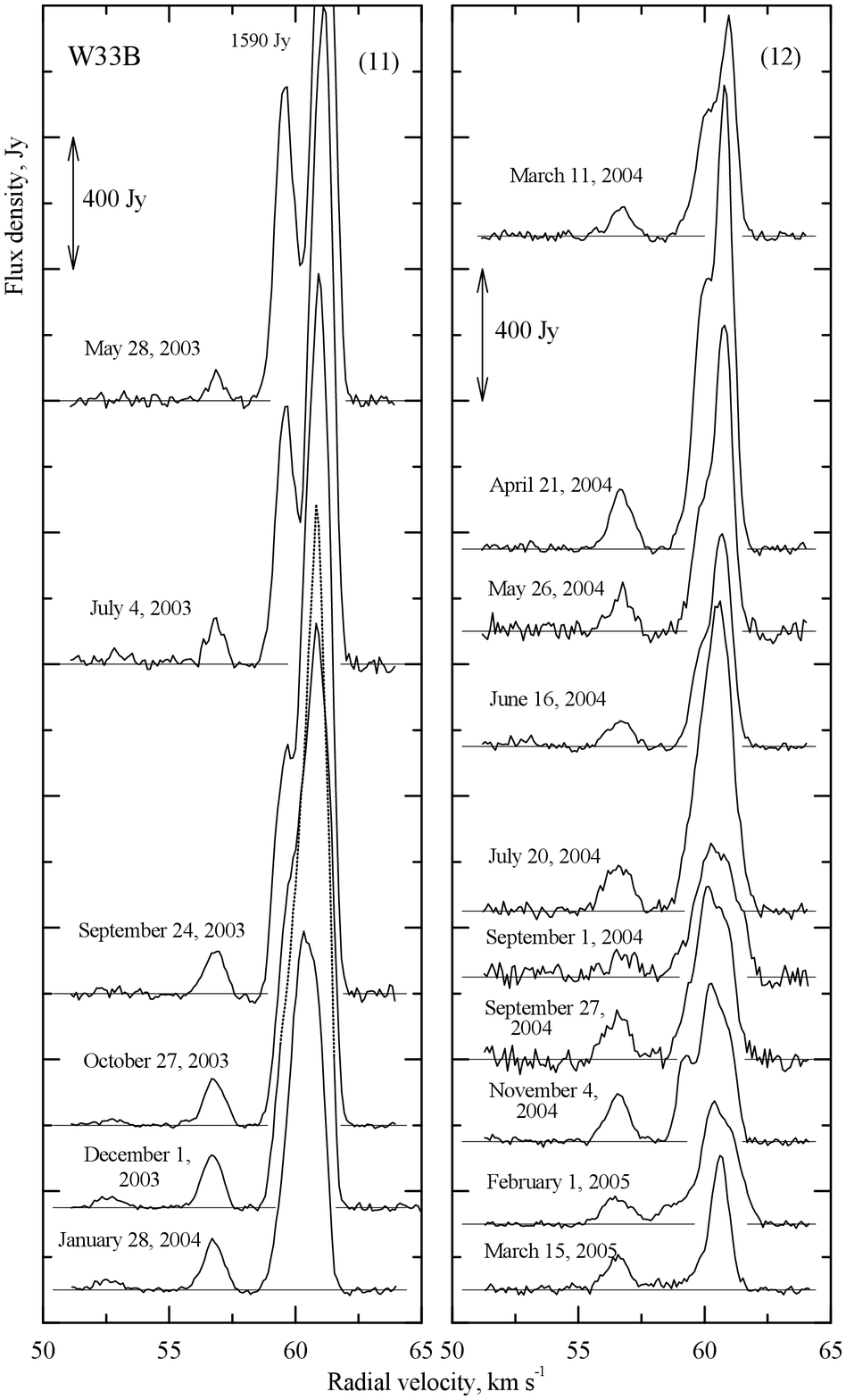} \caption{\small  Continued.} \label{fig7f}
\end{figure*}
}

\addtocounter{figure}{-1}
\onlfig{
\begin{figure*}
\centering
    \includegraphics[width=14cm]{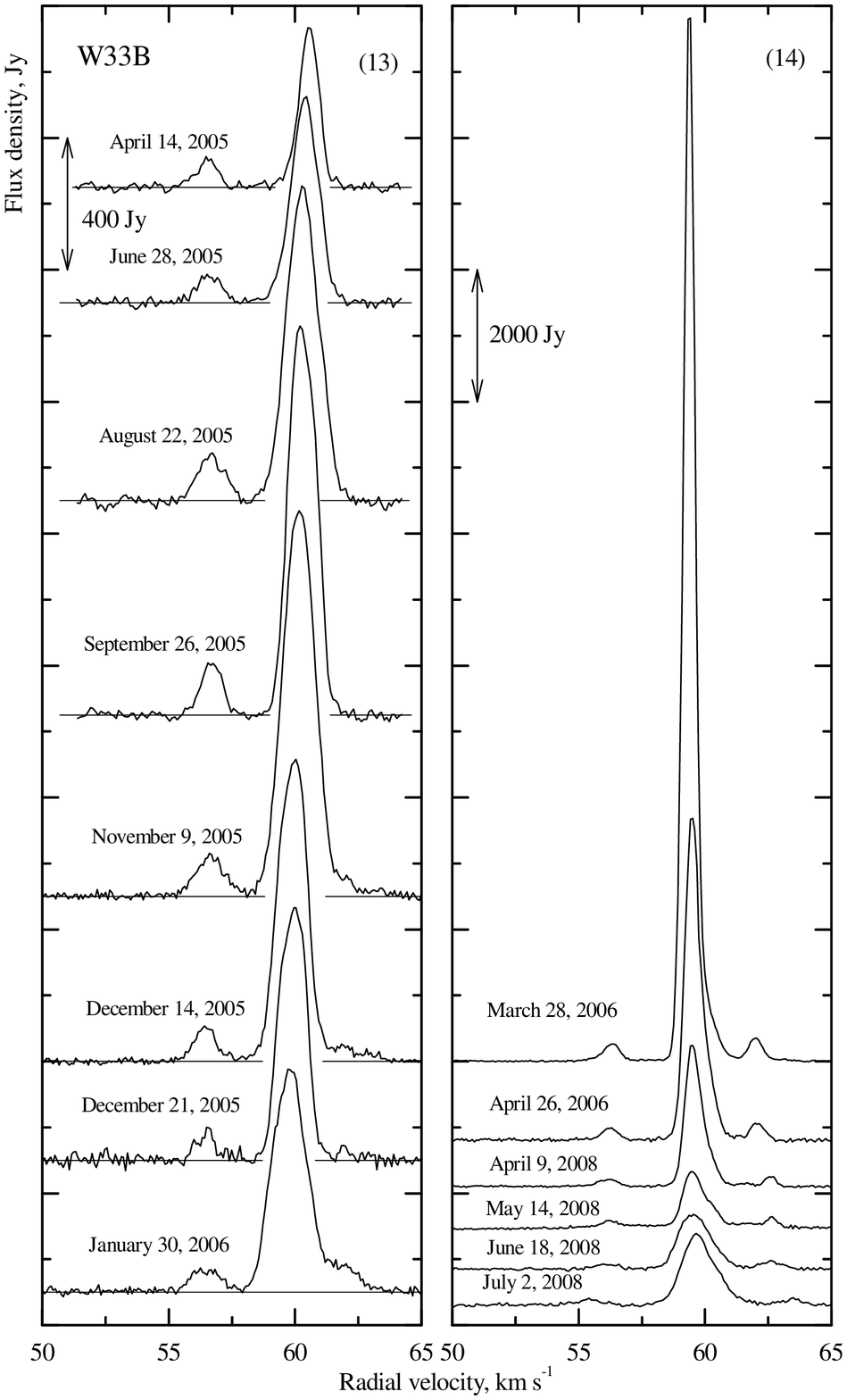} \caption{\small  Continued.} \label{fig7g}
\end{figure*}
}

\addtocounter{figure}{-1}
\onlfig{
\begin{figure*}
\centering
    \includegraphics[width=14cm]{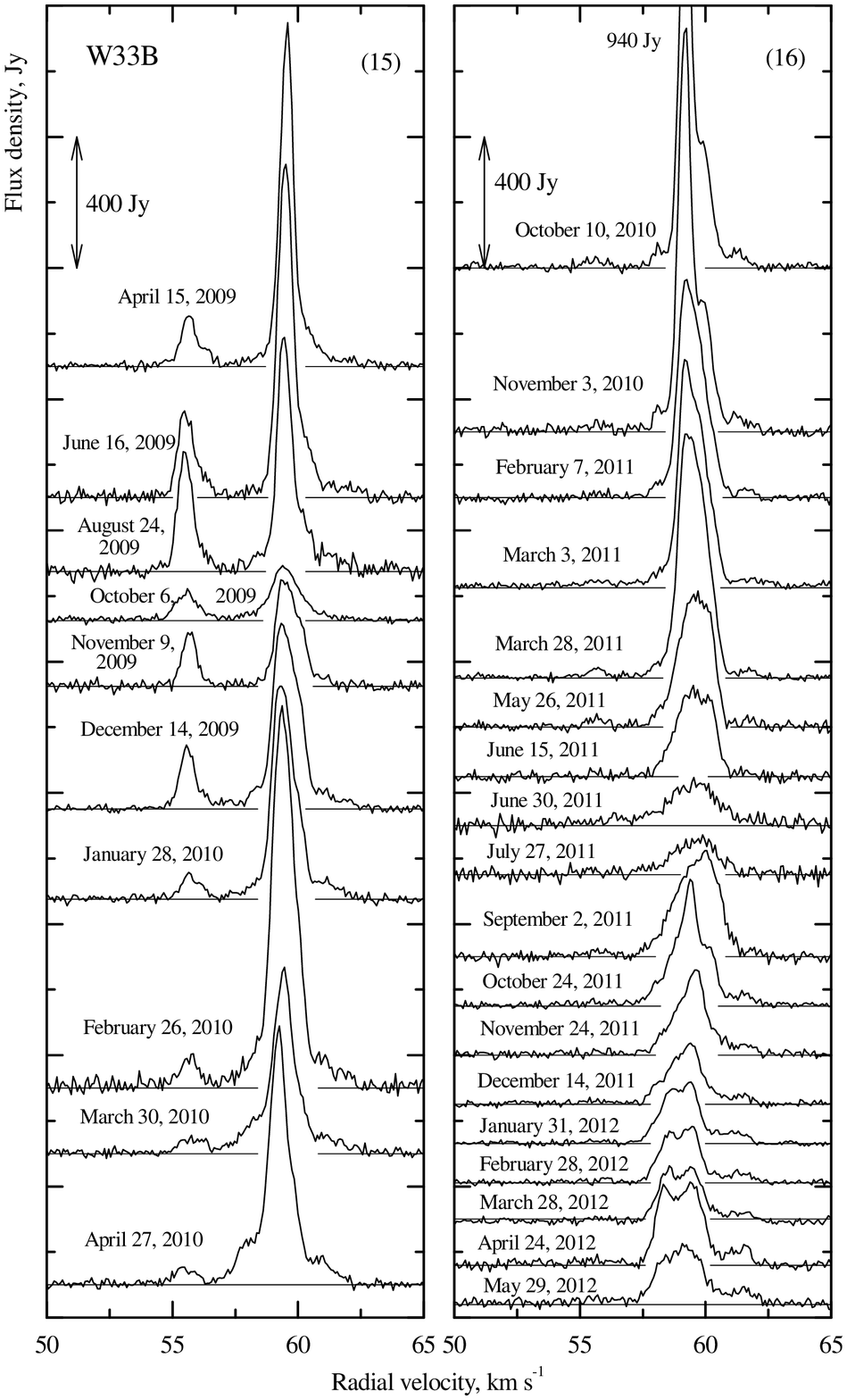} \caption{\small Continued.} \label{fig7h}
\end{figure*}
}

\addtocounter{figure}{-1}
\onlfig{
\begin{figure*}
\centering
    \includegraphics[width=7cm]{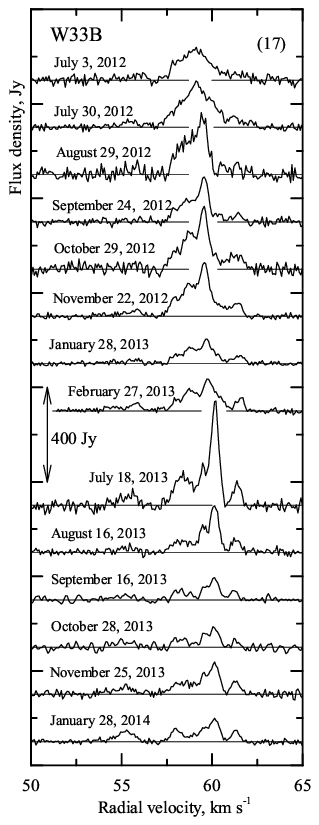} \caption{\small Continued.} \label{fig7i}
\end{figure*}
}

\begin{figure}[!h]
\resizebox{\hsize}{!}{\includegraphics{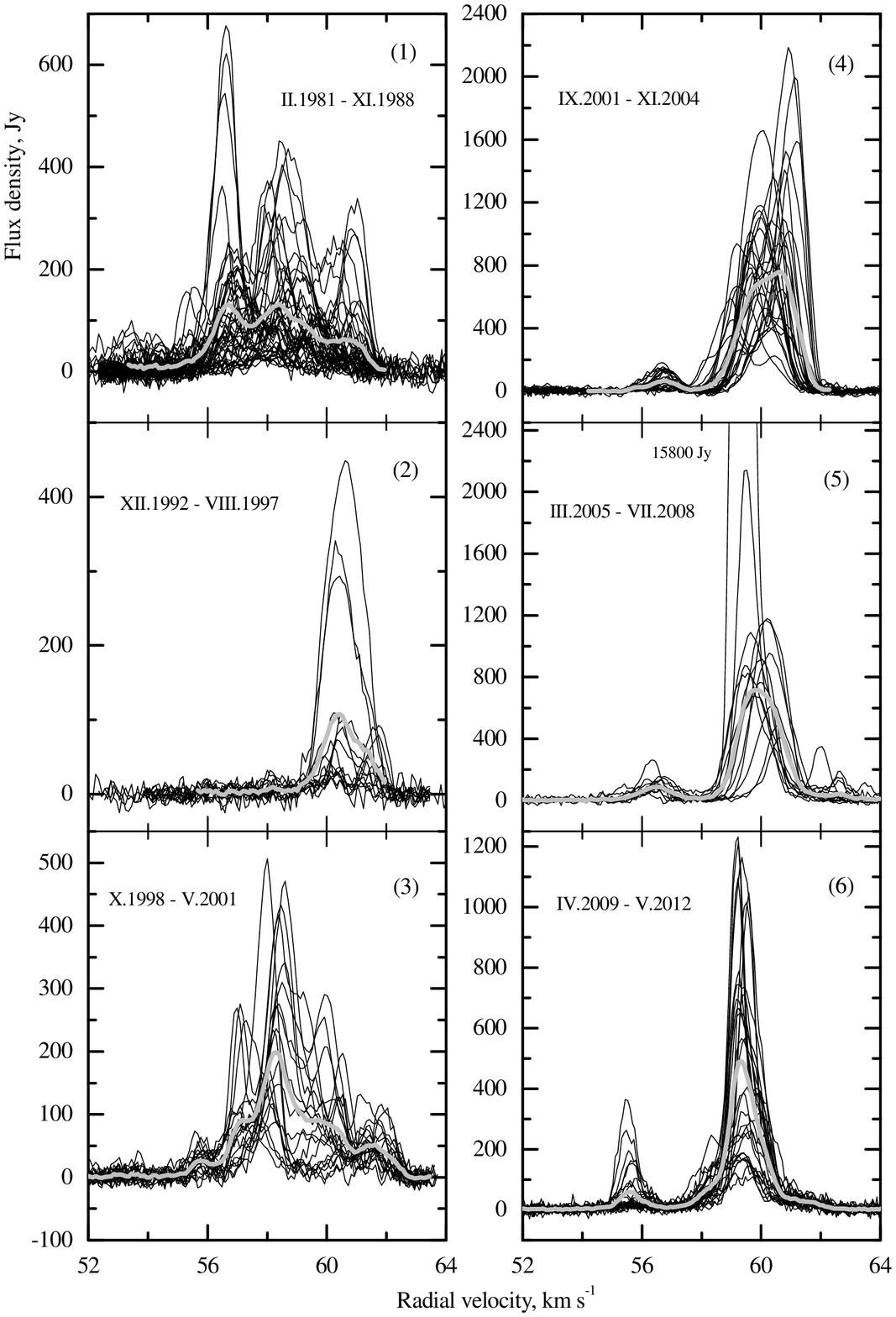}}
\caption{\small Superposition of the H$_2$O 22-GHz spectra for
various time intervals. The separation was done according to the
character of the spectral evolution. Averaged spectra for the
intervals are shown with bold light curves. See text for
details.} \label{fig8}
\end{figure}

\begin{figure}[!h]
\resizebox{\hsize}{!}{\includegraphics{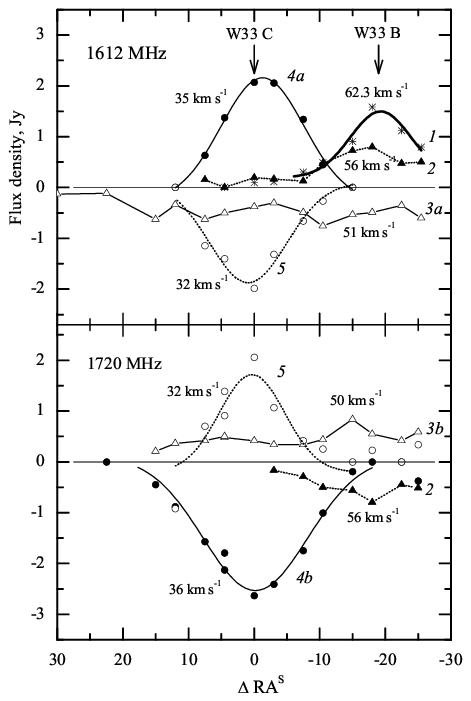}}
\caption{\small Flux density of the features in the satellite OH
lines as a function of right ascension toward W33. $\Delta$\,RA=0
corresponds to the position of W33C. Features with similar radial
velocities are designated \emph{3a, 4a} and \emph{3b, 4b} in the
1612- and 1720-MHz lines, respectively. The strong 1612-MHz maser
feature at 62.3~km~s$^{-1}$ (shown with bold line in the upper
panel) certainly belongs to W33B.}
\label{fig9}
\end{figure}

\begin{figure}[!h]
\resizebox{\hsize}{!}{\includegraphics{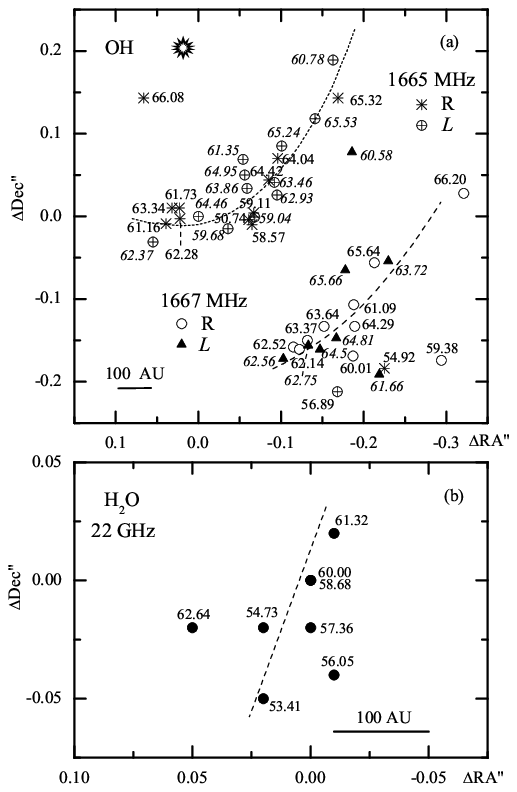}}
\caption{\small (a) VLA map of main-line OH masers for the epoch
1991 \citepads{2000ApJS..129..159A}. 
Right- and left-hand polarised features are shown with different
symbols. Spots' radial velocities are indicated, those of
left-hand polarised features are in \emph{italics}. An asterisk
marks the presumed location of the central star for the OH maser
source based on the arc shape. \protect\\
(b) Positions of VLA H$_2$O maser spots in June 1984
\citepads{1999A&AS..137...43F} 
On both maps the 100~AU bar corresponds to the trigonometric distance
of 2.4~kpc \citepads{2013A&A...553A.117I}. 
Spots' radial velocities are indicated. Dotted curves show the
preferential direction of the growth of the spots' radial
velocities.} 
\label{fig10}
\end{figure}

The emission at 59.8 and 60.4~km~s$^{-1}$ is variable, and the
degree of polarisation also varies. The cause of this can be
turbulent motions of material in the masering region. The
observed emission components have no strong counterparts either
in the main OH lines or in the 1720-MHz satellite line. Thus, we
may observe a new Type IIb OH maser probably associated with an
infrared star. However, the radial-velocity range
($\sim$1.3~km~s$^{-1}$) is too narrow as compared to typical
values for OH masers in late-type giants and supergiants, from
5~km~s$^{-1}$ up to 40~km~s$^{-1}$. The 1612-MHz emission
features being within the velocity interval of the main-line
emission suggests the same distance and physical association with
the 1665/1667-MHz maser. Very long baseline interferometry (VLBI)
observations would be desirable to resolve this question.

\begin{table}
\caption{\small Flux densities in the right- ($F_{\textrm{R}}$) and left-hand
($F_{\textrm{L}}$)\protect\\
circular polarisations in janskys and the degree of polarisation
$p$ \protect\\ in the 1612-MHz OH line.Typical 1$\sigma$ errors
are about 0.1~Jy.}
\label{tab1}
\centering
\footnotesize
\tabcolsep=1.5mm
\begin{tabular}{c|ccc|ccc|ccc}
\hline
\hline
  &
 \multicolumn{9}{c}{Radial velocity, km~s$^{-1}$}  \\
\cline{2-10}
  Date & \multicolumn{3}{c|}{59.1} & \multicolumn{3}{c|}{59.8} & \multicolumn{3}{c}{60.4} \\
\cline{2-10}
   & $F_{\textrm{R}} $ & $F_{\textrm{L}} $ & $p$ & $F_{\textrm{R}} $ & $F_{\textrm{L}} $ & $p$ & $F_{\textrm{R}}$ & $F_{\textrm{L}} $ & $p$  \\
\hline
2012 Mar 8  & 1.4 & 0.3 & 0.65 & 2.7 & 2.6 & 0.02 & 1.0 & \phantom{1}6.9 & 0.75 \\
2012 Apr 7  & 1.6 & 0   &  1   & 3.7 & 2.5 & 0.19 & 0.3 & \phantom{1}7.5 & 0.92 \\
2012 Sept 7 & 1.7 & 0   &  1   & 4.5 & 2.8 & 0.23 & 0.6 & \phantom{1}9.4 & 0.88 \\
2013 May 14 & 1.5 & 0   &  1   & 5.5 & 3.4 & 0.24 & 0.7 &           10.6 & 0.88 \\
2014 Mar 1  & 1.5 & 0   &  1   & 5.1 & 2.6 & 0.32 & 0.3 &           10.7 & 0.95 \\
\hline
\end{tabular}
\end{table}

\subsection{Water-vapour emission}

On the basis of our regular long-term monitoring of W33B
performed with a high spectral resolution, we have studied both
fast and long-term variations of H$_2$O maser emission as well as
the evolution of individual features.

\begin{figure}
\resizebox{\hsize}{!}{\includegraphics{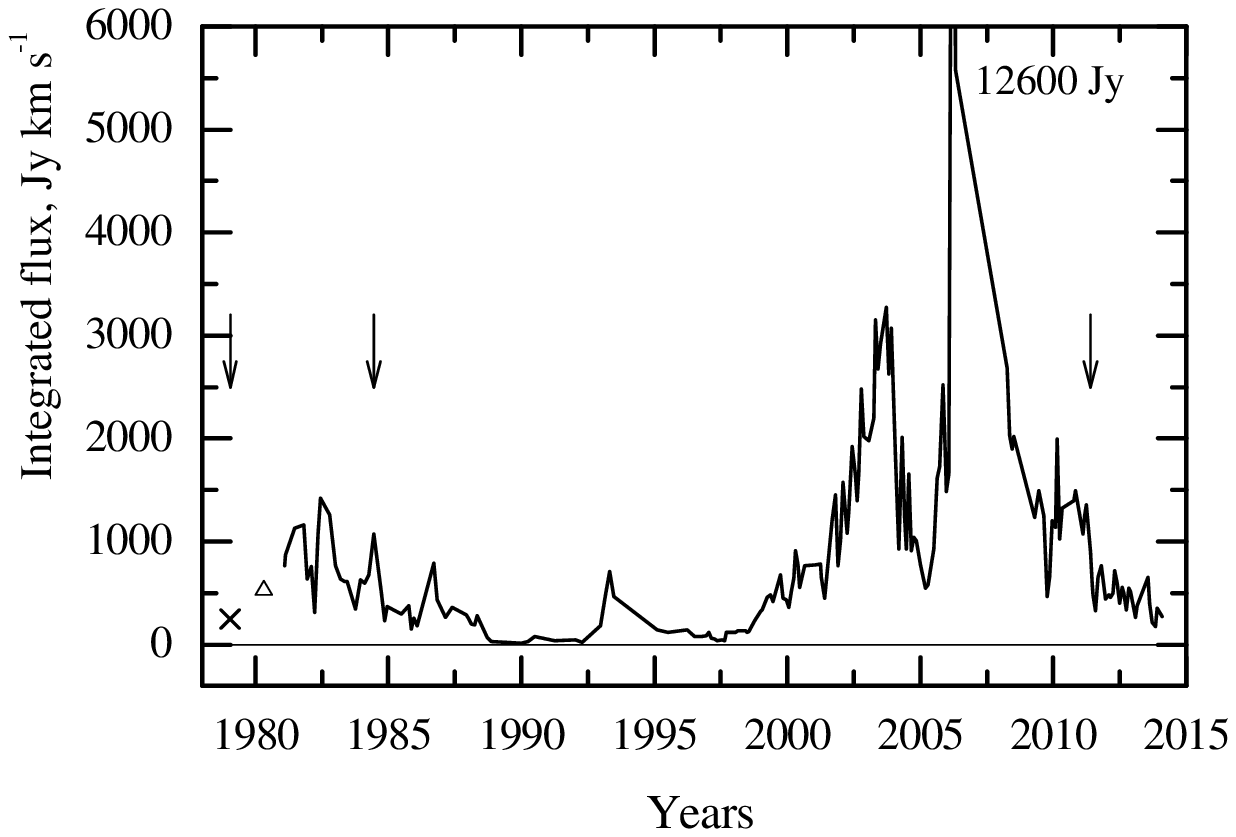}} 
\caption{\small Variations of the integrated H$_2$O line flux. 
Two points at the left taken from observations of other
authors: ($\times$) from \citetads{1981ApJ...243..769L} 
and ($\triangle$) from \citeads{1981ApJ...250..621J}. 
Arrows mark the epochs of the interferometric observations.}
\label{fig11}
\end{figure}

Figure~\ref{fig11} shows variations of the integrated H$_2$O line
flux. The point marked ($\times$) is from
\citetads{1981ApJ...243..769L} 
and ($\triangle$) is from
\citeads{1981ApJ...250..621J}. 
Two stages of maser activity are prominent. Arrows mark the epochs
of the interferometric observations.

We find no cyclicity in the maser variability. We observed two
deep minima: in 1989--1990 and in mid-1997. In addition, we
observed flare-type variability.

There were several interferometric observations of the H$_2$O
maser in W33B. In January 1979 W33B was observed for the first
time by VLBI in the 22-GHz H$_2$O line
by \citetads{1981ApJ...243..769L} 
with a baseline of 845~km. Their map contains four maser features
within a region $\sim 0.05^{\prime\prime}$ i.e. 120--250~AU
depending on the accepted distance, 2.4 or $\sim$5~kpc (see
Introduction).

In June 1984, \citetads{1989A&A...213..339F} 
observed W33B on VLA with a spatial resolution of about
$3\arcsec$ and radial velocity resolution of 1.32~km~s$^{-1}$.
The estimated size of the region hosting maser spots (except for
one) is $0.06\arcsec \times 0.07 \arcsec$ (300$\times$350 AU) in
the right ascension and declination, respectively (see
Figure~\ref{fig10}b). The map centre $\Delta {\rm RA}=0$, $\Delta
{\rm Dec}=0 $ corresponds to RA(2000)=$18^{\rm h}13^{\rm
m}54.7^{\rm s}$, Dec(2000)=$-18\degr 1\arcmin 48.0\arcsec$. The
maser spots are plotted with filled circles. The radial
velocities of each maser spot are indicated. The presence of a
radial-velocity gradient is visible, its direction is shown with
a dashed line. A horizontal bar shows the linear scale of the
W33B region.

\citetads{2013A&A...553A.117I} 
observed H$_2$O masers in the W33 complex on VLBA at nine epochs
between September 2010 and January 2012. In addition to the
trigonometric parallax yielding a distance of 2.4~kpc, they found
proper motions of individual H$_2$O maser features in W33B that
are within 4~mas/year. We note that such proper motions
correspond to projected velocities in the sky plane of
$\sim$45~km~s$^{-1}$, which are by a factor of a few larger than
the maximum line-of-sight velocity spread ($\sim$10~km~s$^{-1}$,
see Figure~\ref{fig8}). This is at variance with the model of
spherically symmetric expansion; the cause of this difference is
not yet clear.

\begin{figure*}
\centering
\includegraphics[width=17cm]{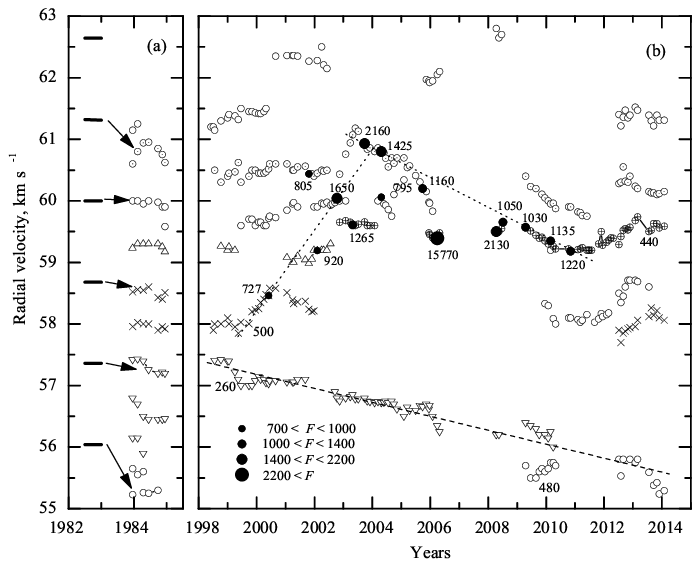}
\caption{\small Radial-velocity variations of the strongest
H$_2$O emission features in W33B. Peaks at different velocities
are shown with different symbols. Main features are shown with
circles of different kinds. Circles of various sizes denote
emission peaks with flux densities that exceeded 700~Jy. The
circle size corresponds to the flux density magnitude. For each
peak its flux density is given in janskys. Dashed and dotted
lines (drawn by eye) present probable radial-velocity drifts of
features persisting throughout our monitoring. Characteristic
timescales are of the order of months.}
\label{fig12}
\end{figure*}

We have identified main spectral components of our monitoring (for
the period June 1984) with the VLA map components. This is shown
with arrows in the left panel of Figure~\ref{fig12}. Since the
spectral resolution of our monitoring is considerably higher, we
have identified and traced the evolution of a larger number of
H$_2$O emission features than the VLA maps. Thus, each feature on
the VLA map may actually correspond to a cluster of several maser
spots. The $V_{\mathrm{LSR}}$ gradient testifies that the
clusters of maser spots form a large-scale organised structure.

Since 2000 the H$_2$O maser was quite active in the
radial-velocity interval 58--62~km~s$^{-1}$, with an intermediate
minimum in early 2005. Since mid-2011 the activity has declined.
In the quoted activity interval we observed flares of individual
features as well as of groups of features. The strongest flare
took place in 2006. In March the flux density reached almost
16,000~Jy.

Figure~\ref{fig12} (right-hand panel) shows radial-velocity
variations of the strongest emission features in W33B. For each of
them circles of different sizes mark the emission maxima in which
the flux density exceeded 700~Jy. The circle size corresponds to
the flux density magnitude. For all maxima flux densities are
given in janskys. In 2000--2003 the emission maximum was moving
toward higher velocities and then in the opposite direction
(dashed curves). This may be due to an ordered arrangement of
maser condensations such as clusters, fragments of shells, etc.
This result is consistent with the H$_2$O structure
on the VLA map \citepads{1999A&AS..137...43F} 
as well as with the map of \citetads{2013A&A...553A.117I}. 

The strongest flare at 59.4~km~s$^{-1}$ in the beginning of 2006
was preceded by complicated variations in the spectrum structure
and by a shift of the emission peak from 60.6 to 59.4~km~s$^{-1}$.
At the epoch of the maximum the line profile was gaussian. The
line was symmetric, its width at half-maximum was
0.53~km~s$^{-1}$. It was local, and it might be associated with a
maser spot. In the presence of turbulent or chaotic motions of
maser condensations in a cluster, two clumps of material can become
superposed in the line of sight, which results in an increase in
the optical depth $\tau$ of the medium. For instance, in the case
of the unsaturated maser an increase in $\tau$ by a factor of 2.3
results in an intensity growth by an order of magnitude.

In some time intervals we observed appreciable radial-velocity
variations of the H$_2$O emission features. Most likely, each
feature is identified not with an individual maser spot but with
some structure, for instance, a filament, a chain with a
radial-velocity gradient or with a more complex formation; see
e.g. \citetads{2003ApJ...598L.115T}, 
\citetads{2007ARep...51..967L}. 
We observed the emission from such structures during intervals of
high activity of the maser source, often in the periods of flare
activity. During propagation in the masering medium of a shock
wave driven by the stellar wind or molecular outflow, regions
with different radial velocities are consecutively excited. This
results in the spectral and spatial drift of the observed
emission peak. In addition, we observed an appreciable
radial-velocity drift of the emission for the cluster of maser
spots from 57.5 to 55.5~km~s$^{-1}$ (dashed line in
Figure~\ref{fig12}b). The drift is indeed considerable to take
into account that the radial-velocity dispersion of all maser
spots in W33B does not exceed 8~km~s$^{-1}$. The drift time
interval falls on the stage of maximum H$_2$O maser activity in
W33B. This is confirmed by the proper motions of the maser
condensations found
by \citetads{2013A&A...553A.117I}. 

The complicated character of the radial-velocity variations may
reflect the presence of turbulent motions of material within an
H$_2$O maser condensation as well as on the scale of compact
clusters of maser spots.

\section{Summary}
We list the main results of our long-term observations of the
hydroxyl maser and of the 30-year monitoring of the water-vapour
maser in the source W33\,B.

\begin{enumerate}
\item We have observed strong variability of the emission features
in the main OH lines 1665 and 1667~MHz.
\item We have detected Zeeman splitting $\sigma$-components in
    the 1665-MHz OH line at 62~km~s$^{-1}$ and in the 1667-MHz
    line at 62 and 64~km~s$^{-1}$. From the amount of splitting
    we have estimated the line-of-sight component of the
    magnetic field for each of the masering regions as 2.0--2.2
    and 3.1~mG, respectively.
\item The profiles of the satellite lines at 1612 and 1720~MHz
    mirror each other. This suggests pumping of the levels of
    these transitions by infrared emission of the source embedded
    in the magnetised interstellar cloud around the maser.
\item We have detected narrowband, strongly variable emission
    in the 1612-MHz line with a high degree of circular
    polarisation, which belongs to a pointlike source.
\item We present an atlas of the H$_2$O $\lambda = 1.35$~cm
    emission spectra toward W33\,B for the time interval from
    November 1981 to May 2006 and from December 2007 to January
    2014. The mean interval between consecutive observational
    sessions was 2.2~months. The radial-velocity resolution was
    0.101~km~s$^{-1}$, and since the end of 2005 it was
    0.0822~km~s$^{-1}$.
\item We detected flares of individual H$_2$O spectral features
    and of groups of features (clusters). The emission features
    probably form filaments, chains with a radial-velocity
    gradient, or more complicated structures including
    large-scale ones.
\item The characteristic variations of OH and H$_2$O maser
    emission suggest the existence of turbulent motions of
    material in the regions of the maser spots' localisation.
\item We have observed two stages of activity of the H$_2$O
    maser with an interval between the main maxima of about
    20~years.
\item The arc-like arrangement of the OH maser spots and the
    large separation between the OH and H$_2$O maser sources
    allow us to suppose that the hydroxyl and water vapour
    masers have independent energy sources.
\end{enumerate}

\begin{acknowledgements}
The Nan\c{c}ay Radio Observatory is the Unit\'e Scientifique de
Nan\c{c}ay of the Observatoire de Paris, associated with the CNRS.
The Nan\c{c}ay Observatory acknowledges the financial support of
the R\'egion Centre in France. The 22~m Pushchino radio telescope
is supported by the Ministry of Science and Education of the
Russian Federation (facility registration number 01-10). This work
was supported by the Russian Foundation for Basic Research
(project code 09-02-00963-a). The authors are grateful to the
staff of the Nan\c{c}ay and Pushchino Radio Astronomy
Observatories for their help with the observations and to the
anonymous referee for useful comments that helped to improve the
paper. This research has made use of the SIMBAD database, operated
at CDS, Strasbourg, France
\end{acknowledgements}
\nocite{*}
\bibliographystyle{aa} 
\bibliography{aa23083-13}
\listofobjects

\label{LastPage}

\end{document}